\documentclass{scrartcl}
\usepackage{amsmath}
\usepackage{amsthm}
\usepackage[numbers]{natbib}
\usepackage{color}
\usepackage{graphicx}
\usepackage{amssymb}

\newtheorem{proposition}{Proposition}
\newtheorem{theorem}[proposition]{Theorem}
\usepackage{bm}
\newcommand{\reals}{\mathbb{R}}
\newcommand\numberthis{\addtocounter{equation}{1}\tag{\theequation}}
\newcommand{\set}[1]{\left\{#1\right\}}
\newcommand{\gp}{\mathcal{GP}}
\newcommand{\norm}[1]{\left\|#1\right\|}

\renewcommand{\complement}{\mathsf{c}}
\newcommand{\wrt}{\textrm{d}}

\title{Probabilistic Numerical Methods for PDE-constrained Bayesian Inverse Problems}

\author{Jon Cockayne\thanks{University of Warwick, j.cockayne@warwick.ac.uk} \and Chris Oates\thanks{University of Technology Sydney, chris.oates@ncl.ac.uk} \and Tim Sullivan \thanks{Free University of Berlin and Zuse Institute Berlin, sullivan@zib.de} \and Mark Girolami\thanks{Imperial College London and Alan Turing Institute, m.girolami@imperial.ac.uk}}

\begin{document}  
\maketitle

\begin{abstract}
	This paper develops meshless methods for probabilistically describing discretisation error in the numerical solution of partial differential equations. This construction enables the solution of Bayesian inverse problems while accounting for the impact of the discretisation of the forward problem.  In particular, this drives statistical inferences to be more conservative in the presence of significant solver error. Theoretical results are presented describing rates of convergence for the posteriors in both the forward and inverse problems. This method is tested on a challenging inverse problem with a nonlinear forward model.
\end{abstract}
 
\section{Introduction} 

Partial differential equations (PDEs) are challenging problems which often have no analytical solution and must be solved numerically. In the style of Probabilistic Numerics (PN) \citep{Hennig:2015jf}, in this work we describe methods for probabilistically modelling the uncertainty in the true solution arising from the numerical approximation. This uncertainty can be thought of as arising from finite computation, as formalised in the Information Complexity literature \cite{Wozniakowski:2009vj}; in solving a problem numerically, we are forced to discretise some aspect of it. In the present work we model this uncertainty as arising from taking a finite number of evaluations of the forcing terms of the system of PDEs.

One of the core principles of probabilistic numerics is that, in complex procedures in which multiple numerical approximations must be composed to produce a final result, the uncertainty from each procedure can combine in a nontrivial way which can lead to incorrect inferences. The example we take here is that of PDE constrained Bayesian inverse problems, in which we wish to estimate parameters of a PDE model in a Bayesian framework, based on observations of a system which is believed to be described by the underlying PDE. In such problems it has been shown that employing an inaccurate PDE solver in the sampling can lead to incorrect inferences in the inverse problem \citep{Conrad:2016gv}. 

There has been recent interest in construction of probabilistic solvers for PDEs. Work by \cite{Conrad:2016gv} constructs a nonparametric posterior distribution for ODEs and PDEs by injecting noise into standard numerical solvers in such a way as to maintain the convergence properties of these solvers. In \cite{Owhadi2015}, the authors discuss a meshless method which is similar to the method discussed herein by modelling the forcing of the PDE. This is developed in \cite{Owhadi:2015vo}, which discusses a methodology for probabilistic solution of PDEs by an hierarchical game-theoretic argument. These latter two approaches do not examine application to inverse problems, however. 

Work from \cite{Cialenco2011} discusses the interpretation of symmetric collocation as the mean function of a Gaussian process prior after conditioning on observed values of the forcing, but applies this methodology predominantly to stochastic differential equations.

\subsection{Structure of the Paper}

We begin by introducing the concept of a probabilistic meshless method and giving some theoretical results related to it. We then show how the posterior measure over the forward solution of the PDE can be propagated to the posterior measure over parameters in a Bayesian inverse problem. Finally we present some numerical results for a challenging nonlinear inverse problem given by the steady-state Allen--Cahn equations.

Proofs for the presented theorems are omitted, and can be found in \cite{Cockayne:2016ts}.

\section{The Probabilistic Meshless Method} \label{sec:pmm}

We now introduce the concept of a probabilistic meshless method (PMM). Consider an open, bounded subset $D$ of $\reals^d$ with Lipschitz boundary $\partial D$. We seek a solution $u\in H(D)$, some Hilbert space of functions defined over $D$, of the following system of operator equations
\begin{alignat*}{2}
	\mathcal{A} u(\bm{x}) = g(\bm{x}) & \quad & \bm{x}\in D \\
	\mathcal{B} u(\bm{x}) = b(\bm{x}) & \quad & \bm{x}\in \partial D . \numberthis \label{eq:pde_system}
\end{alignat*}
Here $\mathcal{A} : H(D) \to H_\mathcal{A}(D)$ and $\mathcal{B} : H(D) \to H_\mathcal{B}(D)$ with $g \in H_\mathcal{A}(D)$ and $b \in H_\mathcal{B}(D)$. $\mathcal{A}$ is associated with a partial differential operator and $\mathcal{B}$ is associated with the boundary conditions of the system. For notational simplicity we restrict attention to systems of two operators, however the methods discussed can be generalised to an arbitrary number of  operator equations.

We proceed in a Bayesian setting by placing a prior measure $\Pi_u$ on $u$, and determining its posterior distribution based on a finite number of observations of the system given in Eq.~\ref{eq:pde_system}. In this work we focus on the most direct observations of said system; namely, we choose sets of design points $\{\bm{x}_{i, \mathcal{A}}\} = X_0^\mathcal{A} \subset D$, $\{\bm{x}_{j, \mathcal{B}}\} = X_0^\mathcal{B} \subset \partial D$ for $i=1,\dots,m_\mathcal{A}$, $j = 1,\dots,m_\mathcal{B}$. We then evaluate the right-hand-side corresponding to each of the operators in the system at these points; $\bm{g} = [g(\bm{x}_{i, \mathcal{A}})]$, $\bm{b} = [b(\bm{x}_{i, \mathcal{B}})]$.

It remains to specify our prior distribution. Here we choose a Gaussian process prior $\Pi_u = \gp(m, k)$. Recall that a Gaussian Process is characterised by its mean function $m$ and its covariance function $k$, and the property that, if $u\sim \gp(m, k$) then for any set of points $\set{\bm{x}_i} \subset \reals^d$, $i=1,\dots,n$
\begin{align*}
	u(X) &\sim \mathcal{N}(\bm{\mu}, \Sigma) \\
	[\bm{\mu}]_i &= m(\bm{x}_i) \\
	[\Sigma]_{ij} &= k(\bm{x}_i, \bm{x}_j)
\end{align*}
As is common in the literature we will use a centred Gaussian process prior; $\Pi_u = \gp(0, k)$. Define
\begin{equation*}
	\mathcal{L} = \begin{bmatrix} \mathcal{A} \\ \mathcal{B} \end{bmatrix} \quad 
	\bar{\mathcal{L}} = \begin{bmatrix} \bar{\mathcal{A}} & \bar{\mathcal{B}} \end{bmatrix}
\end{equation*}
and furthermore for sets $X=\{x_i\}$, $i=1,\dots,N$, $Y=\{y_j\}$, $j=1,\dots,M$ let $K(X, Y)$  denote the Gram matrix of $K$ applied to $X$ and $Y$; $[K(X, Y)]_{ij} = k(x_i, y_j)$. Similarly $[\mathcal{A}K(X, Y)]_{ij} = \mathcal{A}k(x_i, y_j)$, etc. Then
\begin{align*}
	\mathcal{L}\bar{\mathcal{L}} K(X_0, X_0) &= \begin{bmatrix}
		\mathcal{A}\bar{\mathcal{A}} K(X_0^\mathcal{A}, X_0^\mathcal{A}) 
		& \mathcal{A}\bar{\mathcal{B}} K(X_0^\mathcal{A}, X_0^\mathcal{B}) \\
		\mathcal{B}\bar{\mathcal{A}} K(X_0^\mathcal{B}, X_0^\mathcal{A}) 
		& \mathcal{B}\bar{\mathcal{B}} K(X_0^\mathcal{B}, X_0^\mathcal{B})
	\end{bmatrix} \\
	\mathcal{L} K(X_0, X) &= \begin{bmatrix} 
		\mathcal{A} K(X_0^\mathcal{A}, X) \\
		\mathcal{B} K(X_0^\mathcal{B}, X)
	\end{bmatrix} \\
	\bar{\mathcal{L}} K(X, X_0) &= \begin{bmatrix}
		\bar{\mathcal{A}} K(X, X_0^\mathcal{A}) & \bar{\mathcal{B}} K(X, X_0^\mathcal{B}) 
	\end{bmatrix}
\end{align*}
Here $X$ is to be interpreted as a set of points at which we evaluate those functions drawn from the posterior distribution, in contrast with $X_0 = X_0^\mathcal{A} \cup X_0^\mathcal{B}$ which is the set of points at which evaluations of the forcing terms are taken.

\begin{proposition}[Probabilistic Meshless Method] \label{prop:pmm}

	Assume $\mathcal{A}$ and $\mathcal{B}$ are linear operators. Then the posterior distribution $\Pi_u^{\bm{g}, \bm{b}}$ over the solution of the PDE, conditional on the data $\bm{g}, \bm{b}$ is such that, for $u\sim \Pi_u^{\bm{g}, \bm{b}}$ we have
	\begin{align*}
		u(X) &\sim \mathcal{N}(\bm{\mu}, \Sigma) \\
		\bm{\mu} &= \bar{\mathcal{L}} K(X, X_0) \left[ \mathcal{L} \bar{\mathcal{L}}K(X_0, X_0)\right]^{-1} \begin{bmatrix} \bm{g}^\top & \bm{b}^\top \end{bmatrix}^\top  \numberthis \label{eq:posterior_mean}\\
		\Sigma &= K(X, X) - \bar{\mathcal{L}} K(X, X_0) \left[ \mathcal{L} \bar{\mathcal{L}}K(X_0, X_0)\right]^{-1} 
		\mathcal{L} K(X_0, X)
	\end{align*}
\end{proposition}
\noindent
Note that the mean function in Eq.~\ref{eq:posterior_mean} is the same as the numerical solution to the PDE that would be obtained using the method of symmetric collocation \citep{Fasshauer1999}.

Thus far we not discussed the choice of prior covariance $k$. There are several interesting choices in the literature. Work in \cite{Owhadi2015} proposes use of a covariance which encodes information about the system through its Green's function; \cite{Cockayne:2016ts} examined the properties of this choice in more detail. However, reliance on the Green's function, which is not in general available in closed-form for complex systems, is a significant drawback. In practice we will generally posit a prior covariance directly by examining the system in question and selecting a prior which encodes a suitable level of differentiability.

We now present a theoretical result describing the rate of convergence of the posterior measure $\Pi_u^{\bm{g}, \bm{b}}$. Denote by $\rho$ the differential order of the PDE; that is, the maximum number of derivatives of $u$ required. Furthermore denote by $\beta$ the smoothness of the prior; the number of weak derivatives that almost surely exist under the prior measure. Lastly, define $h$ to be the ``fill distance'' of the design points $X_0$:
\begin{equation*}
	h = \sup_{\bm{x}\in D} \min_{\bm{x}' \in X_0} \norm{\bm{x}-\bm{x}'}_2
\end{equation*}

\begin{theorem}[Rate of Convergence]
	For a ball $B_\epsilon(u_0)$ of radius $\epsilon$ centred on the true solution $u_0$ of the system \eqref{eq:pde_system}:

	\begin{equation*}
		\Pi_u^{\bm{g}, \bm{b}}(B_\epsilon(u_0)^\complement) = \mathcal{O}\left(\frac{h^{2\beta - 2\rho - d}}{\epsilon}\right)
	\end{equation*}
	where $\complement$ denotes the set complement.
\end{theorem}

\subsection{Illustrative Example: The Forward Problem} \label{sec:illustrative_fwd}
We conclude this section by examining the performance of the probabilistic meshless method for a simple 1-dimensional PDE. Consider the system
\begin{alignat*}{2}
	-\nabla^2 u(x) &= \sin(2\pi x) & \quad & x \in (0,1) \\
	u(x) &= 0 & \quad & x =0,1
\end{alignat*}
the solution to which can be computed by direct integration to be $u(x) = -(2\pi^2)^{-2} \sin(2\pi x)$. We compute the PMM solution to this PDE with varying number of design points. In this setting the Green's function for the system is available explicitly, and so we used its associated prior covariance as suggested in \cite{Owhadi2015}; full details are available in \cite{Cockayne:2016ts}.

\begin{figure}
		\includegraphics[width=0.45\textwidth]{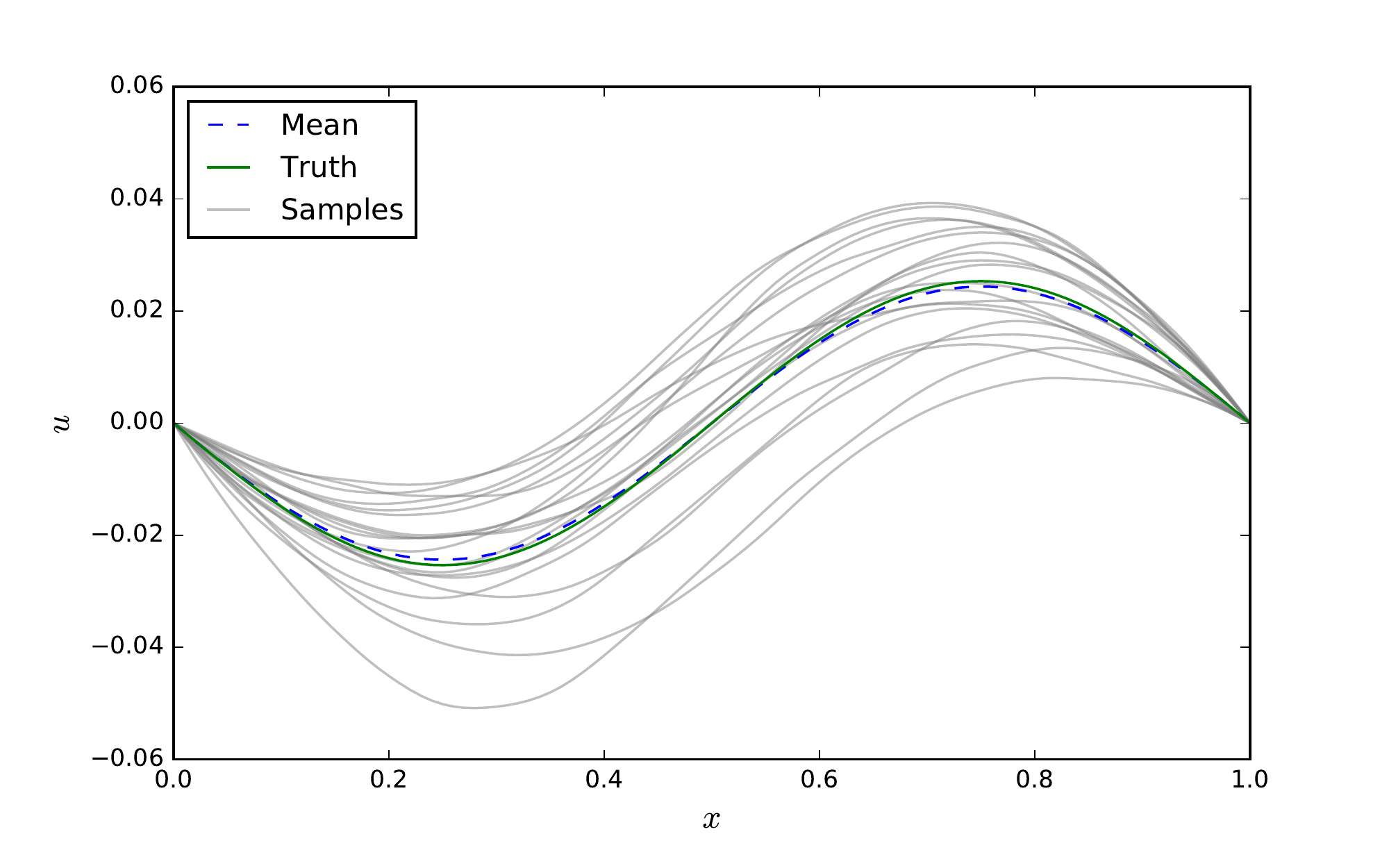}
		\includegraphics[width=0.45\textwidth]{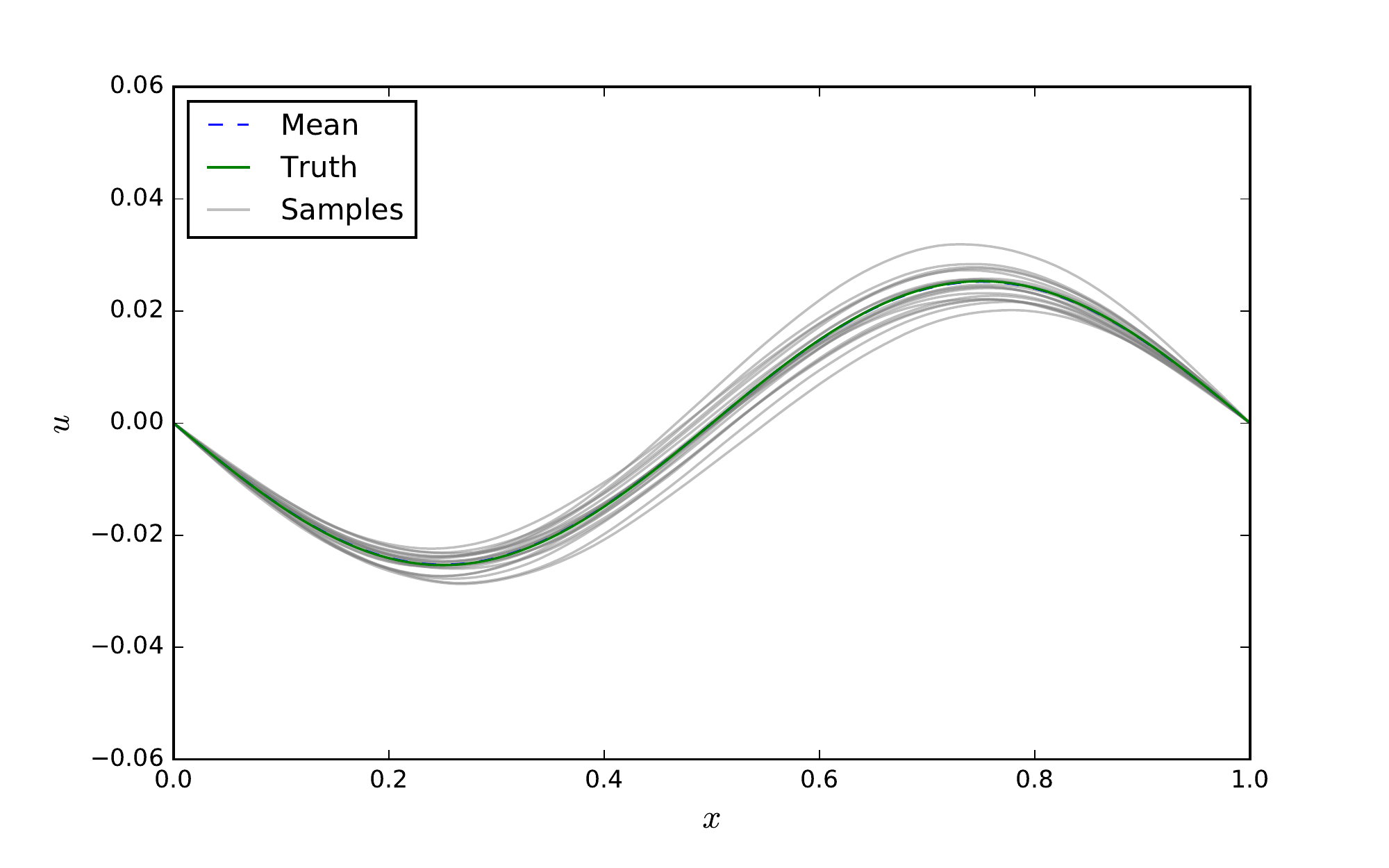}
		\caption{Samples from the posterior distribution over the unkown solution to a one-dimensional PDE, with $m_\mathcal{A} = 10$ (left) and $m_\mathcal{A}=40$ (right).}
		\label{fig:forward_posterior_samples}
	\end{figure}

\begin{figure}
		\includegraphics[width=0.45\textwidth]{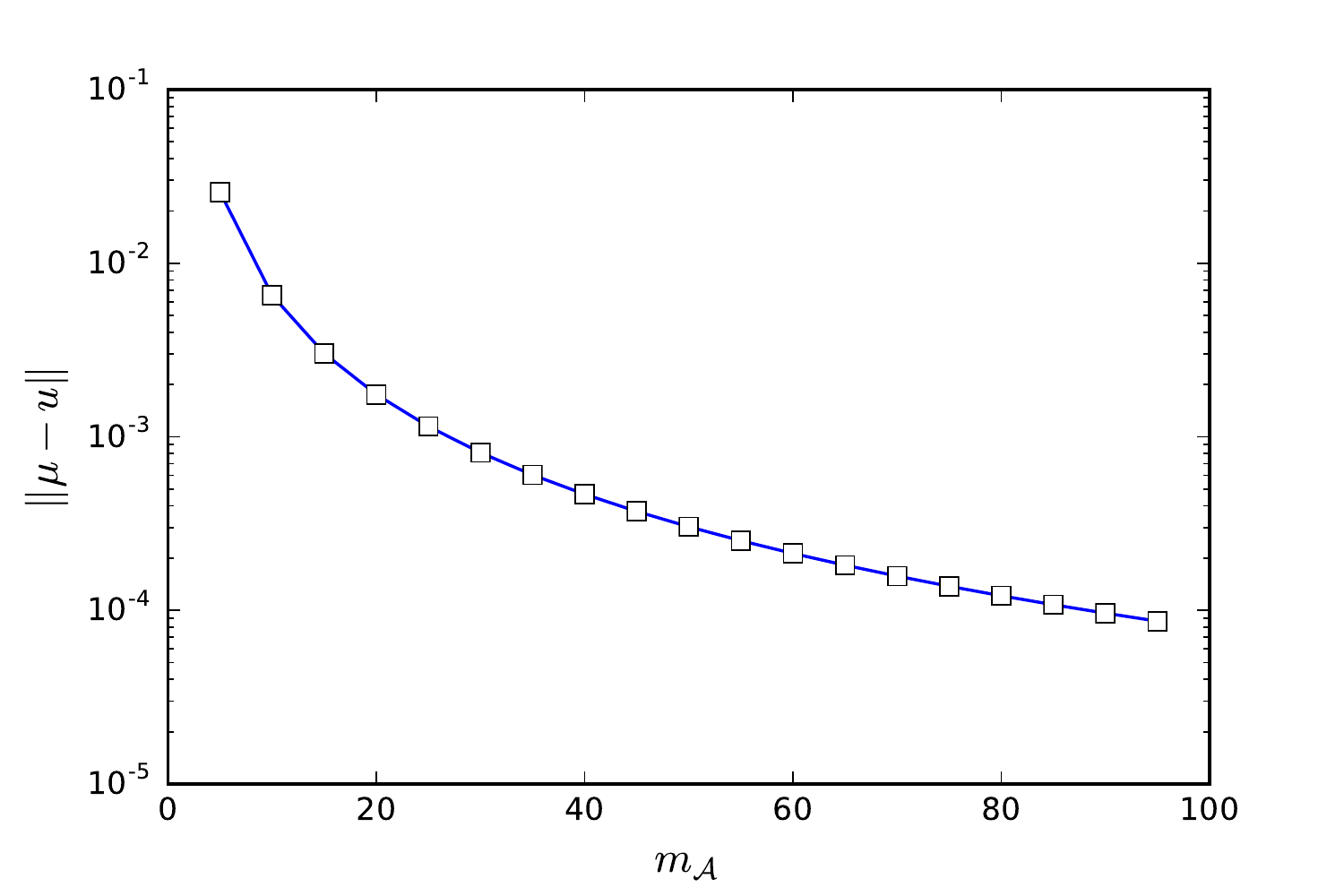}
		\includegraphics[width=0.45\textwidth]{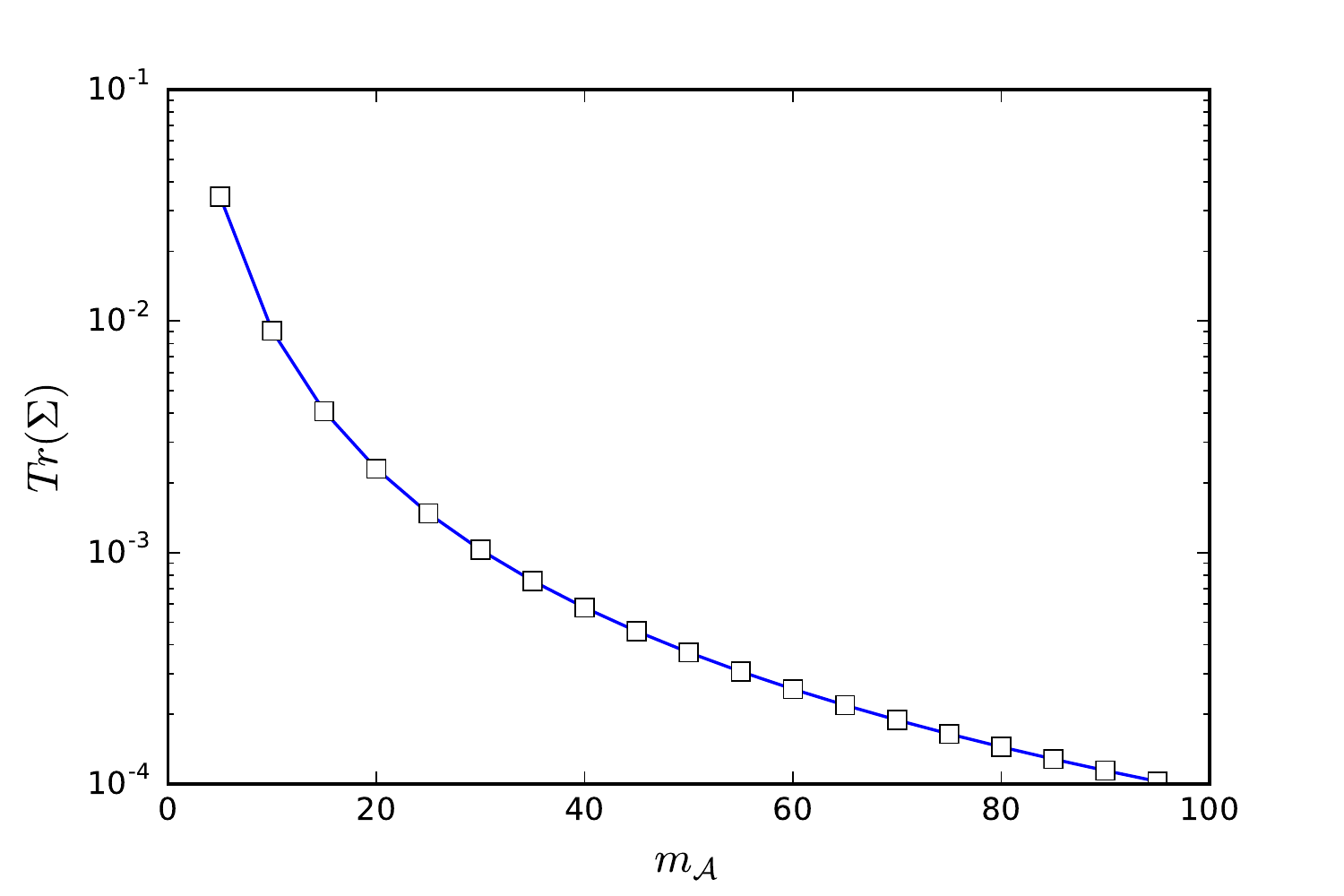}
		\caption{Convergence of mean function (left) and posterior covariance trace (right) as the number of design points $m_\mathcal{A}$ is increased.}
		\label{fig:forward_posterior_convergence}
	\end{figure}

Samples from the posterior distribution can be seen in Fig.~\ref{fig:forward_posterior_samples}; note how, even with 20 design points, there is still significant posterior uncertainty. Convergence plots as the number of design points is increased are shown in Fig.~\ref{fig:forward_posterior_convergence}

\section{Application to Bayesian Inverse Problems} \label{sec:inverse_problems}

We now turn to an examination of how the PMM, constructed in the previous section, can be applied in Bayesian inverse problems. We now have a system in which we assume the operator $\mathcal{A}$ depends upon some parameter $\theta$, which we emphasise in the below system:
\begin{alignat*}{2}
	\mathcal{A}_\theta u(\bm{x}) = g(\bm{x}) & \quad & \bm{x}\in D \\
	\mathcal{B} u(\bm{x}) = b(\bm{x}) & \quad & \bm{x}\in \partial D .
\end{alignat*}

In a Bayesian inverse problem we place a prior distribution over $\theta$, $\theta \sim \Pi_\theta$, and seek to determine its posterior distribution $\Pi_\theta^{\bm{y}}$ based on data $\bm{y}$ collected at locations $\set{\bm{x}_i} \subset D$, $i=1,\dots,n$. Further details on Bayesian inverse problems can be found in \cite{Stuart2010}.

Such a posterior distribution is usually intractable and must be investigated by sampling, which involves solution of the underlying system of PDEs as the sampler visits different values of $\theta$. We assume that the data is obtained by direct observation of the solution $u$ at these locations, corrupted with Gaussian noise
\begin{equation*}
	y_i = u(\bm{x}_i) + \xi_i
\end{equation*}
where $\bm{\xi} \sim \mathcal{N}(0, \Gamma)$. Our likelihood is thus given by
\begin{equation}
	p(\bm{y} | \theta, u) = \mathcal{N}(\bm{y}; \bm{u}, \Gamma) \label{eq:likelihood}
\end{equation}
where $\bm{u}, \bm{y}$ are each vectors in $\reals^n$, with $[\bm{u}]_i = u(\bm{x}_i; \theta)$ and $[\bm{y}]_i = y_i$. 

Since the solution $u$ to the PDE system is inaccessible it is common to replace $u$ with an approximation $\hat{u}$ obtained by some numerical scheme. We instead use the PMM as the forward solver, obtaining a measure $\Pi_u^{\bm{g}, \bm{b}}$ describing our uncertainty. We may them marginalise $u$ in Eq.~\ref{eq:likelihood} over this measure to obtain
\begin{align*}
	p_{\textrm{PN}}(\bm{y} | \theta) &= \int p(\bm{y} | \theta, u) \;\Pi_u^{\bm{g}, \bm{b}}(\wrt u) \\
	&= \mathcal{N}(\bm{y}; \mu(\theta), \Gamma + \Sigma(\theta)) \numberthis \label{eq:pn_likelihood}
\end{align*}
where $\mu(\theta), \Sigma(\theta)$ are as in Prop.~\ref{prop:pmm}, and we have emphasised the dependence on $\theta$. This is thus similar to the standard approach of replacing $u$ with $\hat{u}$ in Eq.~\ref{eq:likelihood}, but we compensate for the inaccuracy of the forward solver with an additive covariance term $\Sigma$ incorporating the uncertainty in the posterior distribution for the forward problem.

We now present a result which guarantees consistency in the inverse problem when we replace the likelihood in Eq.~\ref{eq:likelihood} with that in Eq.~\ref{eq:pn_likelihood}.

\begin{proposition}{(Inverse Problem Consistency)}
	Let $\Pi_{\theta, \textrm{PN}}^{\bm{y}}$ be the posterior distribution which uses the PN likelihood given in Eq.~\ref{eq:pn_likelihood}. Assume that the posterior distribution $\Pi_\theta^{\bm{y}}$ contracts such that $\Pi_\theta^{\bm{y}} \to \delta(\theta_0)$ as $n \to \infty$, a Dirac measure centred on the true value of $\theta$, $\theta_0$. Then $\Pi_{\theta, \textrm{PN}}^{\bm{y}}$ contracts such that $\Pi_{\theta, \textrm{PN}}^{\bm{y}} \to \delta(\theta_0)$ provided
	\begin{equation*}
		h = o(n^{-1/(\beta - \rho - d/2)})
	\end{equation*}

\end{proposition}

\subsection{Illustrative Example: The Inverse Problem}

We now return to the previous illustrative example to demonstrate the use of a probabilistic solver in the inverse problem. Consideronsider the system
\begin{alignat*}{2}
	-\nabla\cdot \theta \nabla u(x) &= \sin(2\pi x) & \quad & x \in (0,1) \\
	u(x) &= 0 & \quad & x =0,1
\end{alignat*}
with the goal of inferring the parameter $\theta$. Data $y_i$ was generated from the explicit solution to this problem with $\theta = 1$ at locations $x=0.25, 0.75$, and corrupted with Gaussian noise with distribution $\mathcal{N}(0, 0.01^2)$.

\begin{figure}
	\includegraphics[width=\textwidth]{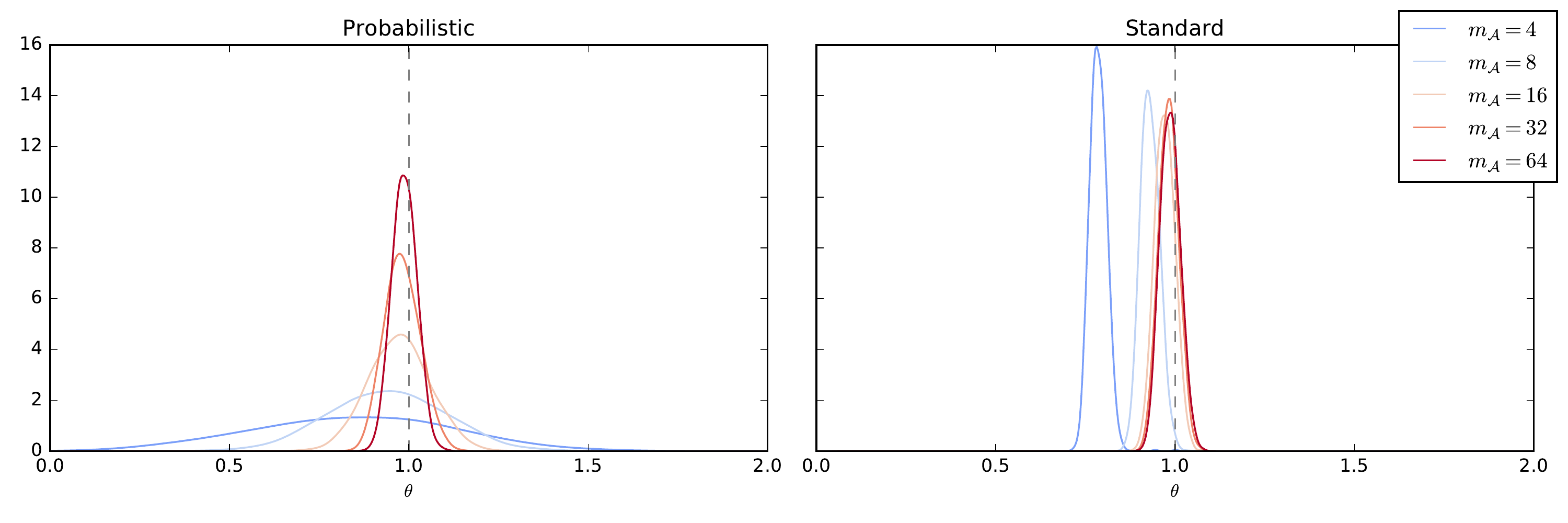}
	\caption{Posterior distributions over $\theta$ with varying numbers of design points, on the left using the PMM, and on the right the standard approach of using a plug-in estimate for the PDE solution, here given by symmetric collocation.}
	\label{fig:inverse_posteriors}
\end{figure}

In Fig.~\ref{fig:inverse_posteriors} we compare posterior distributions for $\theta$ generated with the PMM versus the standard approach of plugging a numerical solution of the PDE into the likelihood and ignoring discretisation error. The numerical method used in the standard approach was symmetric collocation, the most natural comparison. Note that when using collocation the posterior distributions are peaked and biased for small $m_\mathcal{A}$, and that the posterior uncertainty does not appear to depend on the number of design points. Conversely when using the probabilistic method we see that for small $m_\mathcal{A}$ the posterior distributions are wide and flat, while as $m_\mathcal{A}$ increases the distributions peak and centre on the true value of $\theta$. 
Thus, with a standard numerical method the posteriors over $\theta$ do not take into account the quality of the numerical solver used; for poor forward solvers based on coarse discretisations, the posteriors produced are as confident as those produced with a fine, accurate numerical solver. With a probabilistic forward solver the variance in the forward solver is propagated into the inverse problem, resulting in robust inferences even when the discretisation is coarse.

\section{A Nonlinear Example} \label{sec:numerics}

We now present an application of the methods discussed herein to a nonlinear partial differential equation known as the steady-state Allen--Cahn system, a model from mathematical physics describing the motion of boundaries between phases in iron alloys. This is given by
\begin{alignat*}{2}
	-\delta \nabla^2 u + \delta^{-1} (u^3 - u) &= 0 & \quad & \bm{x} \in (0,1)^2 \\
	u &= +1 & \quad & x_1 \in \set{0,1}, x_2 \in (0,1) \\
	u &= -1 & \quad & x_2 \in \set{0,1}, x_1 \in (0,1) \numberthis \label{eq:ac}
\end{alignat*}

\begin{figure}
	\includegraphics[width=\textwidth]{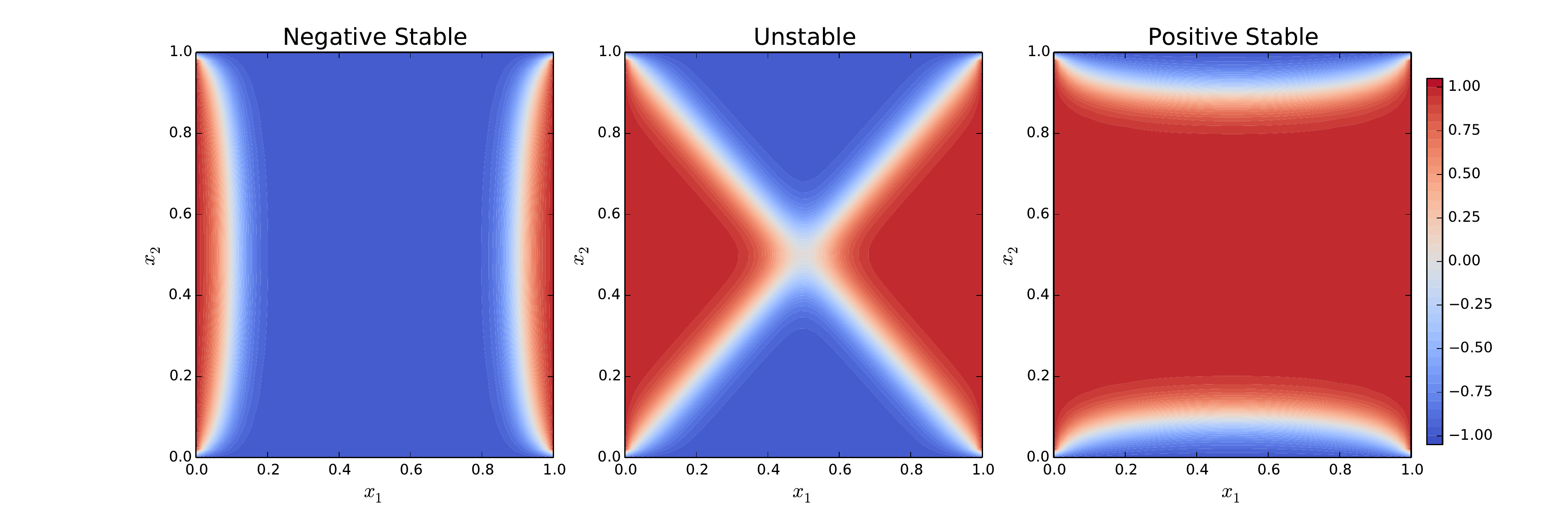}
	\caption{Solutions to the Allen--Cahn system for $\delta = 0.04$}
	\label{fig:allen_cahn_solutions}
\end{figure}

We phrase this as an inverse problem for determining $\delta$. This system is noteworthy for the fact that it does not admit a unique solution; the three solutions to this system for $\delta = 0.04$ are shown in Fig.~\ref{fig:allen_cahn_solutions}. These were generated using the deflation technique described in \cite{Funke2013}.

Since this is a nonlinear system the posterior distribution will not be Gaussian, and we must resort to sampling techniques to explore the posterior distribution. In brief, we introduce a latent function $z$ and rearrange the system as follows:
\begin{align}
	-\delta \nabla^2 u - \delta^{-1} u &= z \label{eq:ac_rearrange_1} \\
	\delta^{-1} u^3 &= -z \label{eq:ac_rearrange_2}
\end{align}
Note that by adding Eq.~\ref{eq:ac_rearrange_1} and Eq.~\ref{eq:ac_rearrange_2} we return to the original equation describing the interior dynamics given in Eq.~\ref{eq:ac}. However Eq.~\ref{eq:ac_rearrange_2} is monotonic and thus invertible; by inverting this we arrive at a new system:
\begin{align*}
	-\delta \nabla^2 u - \delta^{-1} u &= z \\
	u &= (-\delta z)^{1/3}
\end{align*}
This system is equivalent to the original system but, importantly, is \emph{linear}. Thus by the introduction of $z$ we are able to arrive at a new system which can be solved using the PMM.

It remains to describe $z$, a latent function whose value is unknown. We seek to marginalise $z$ in the likelihood
\begin{equation}
	p(\bm{y} | \delta) = \int p(z | \delta) \int p(\bm{y} | u) \;\Pi_u^{\bm{g}, \bm{b}, z}(\wrt u) \; \wrt z \label{eq:intractable_likelihood}
\end{equation}
where $\Pi_u^{\bm{g}, \bm{b}, z}$ is now additionally conditioned on a known value for $z$.
This integral is intractable. However by sampling from the posterior distribution over $\delta$ by pseudo-marginal MCMC it is sufficient to produce an unbiased estimate of this quantity. This is accomplished by importance sampling; we assume an improper prior $p(z | \delta) \propto 1$ and approximate Eq.~\ref{eq:intractable_likelihood} by the Monte-Carlo estimate
\begin{equation*}
	p(\bm{y} | \delta) \approx \frac{1}{M}\sum_{i=1}^M \frac{\int p(\bm{y} | u) \;\Pi_u^{\bm{g}, \bm{b}, z_i}(\wrt u)}{r(z_i | \bm{y}, \delta)}
\end{equation*}
for $z_i \sim r(z | \bm{y}, \delta)$.

The importance distribution $r(z | \bm{y}, \delta)$ is chosen by solving the original system in Eq.~\ref{eq:ac} using the techniques described in \cite{Funke2013}, with a coarse finite-element solver. This gives estimates $\set{\hat{u}_1, \hat{u}_2, \hat{u}_3}$ for the solution given a value of $\delta$. By applying Eq.~\ref{eq:ac_rearrange_2} to these estimates we obtain estimates of three values of $z$; $\set{\hat{z}_1, \hat{z}_2, \hat{z}_3}$.

To handle the multimodality in the solutions we extend the state-space of the inverse problem to include the solution index $j$. The importance distribution is constructed as a Gaussian distribution
\begin{equation*}
	z \sim \gp(\hat{z}_j, k)
\end{equation*}
with $r(z | \bm{y}, \delta, j)$ thus the appropriate multivariate Gaussian density after the field for $z$ has been discretised. Discretisation points are necessarily chosen to match $X_0^\mathcal{A}$, the design points for $u$ in the interior of the domain.

For application of the PMM we choose a squared-exponential prior covariance
\begin{equation*}
	k(x,x') = \exp\left( -\frac{\norm{x-x'}_2^2}{2\ell^2} \right)
\end{equation*}
which is known to describe infinitely-differentiable functions. This choice is motivated by the high differential order required by the PDE; since we must be able to apply both the operator and the adjoint to the kernel, in this case we require that the covariance be twice differentiable in each argument, which amounts to a four-times differentiable covariance if the covariance chosen is isotropic.

The length-scale hyper-parameter $\ell$ was incorporated into the MCMC procedure, endowed with a half-Cauchy hyper-prior as recommended in \cite{Gelman2006}. The parameter of interest $\delta$ was endowed with a uniform prior over the interval $(0.02, 0.15)$, in which the PDE was empirically found to consistently have three solutions.

\begin{figure}
	\includegraphics[width=0.45\textwidth]{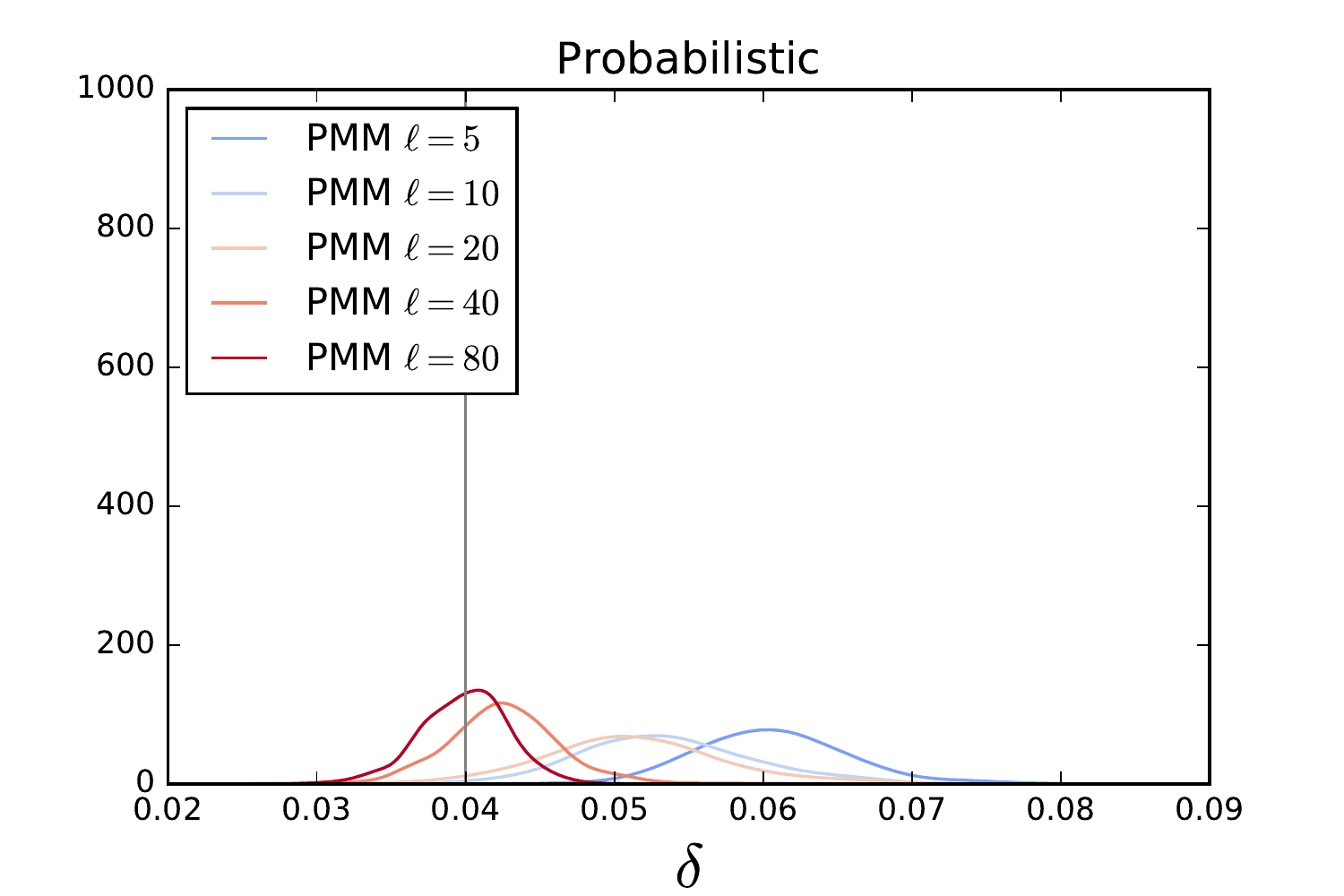}
	\includegraphics[width=0.45\textwidth]{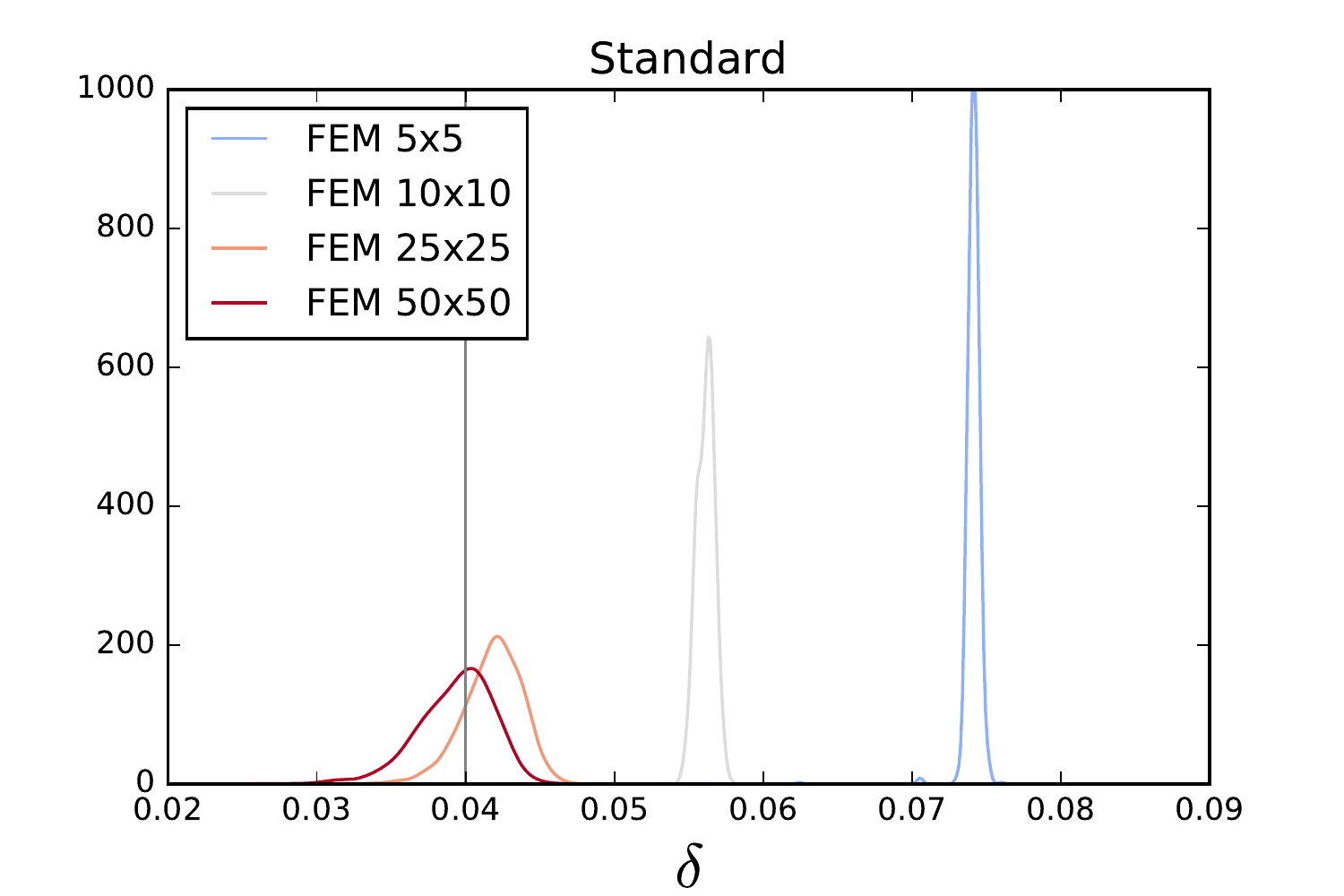}
	\caption{Posterior distributions for $\delta$ obtained by use of the technique described herein (left) versus a standard Finite Element solver that does not model discretisation error (right).}
	\label{fig:ac_posteriors}
\end{figure}

Posterior distributions for $\delta$ generated using this methodology are shown in Fig.~\ref{fig:ac_posteriors}; these are compared with posterior distributions generated using a finite-element forward solver. In the finite-element case we see a more extreme version of the bias shown in Fig.~\ref{fig:inverse_posteriors} for coarse grids, whereas when using a probabilistic forward solver the posteriors are once again wider to account for an inaccurate forward solver.

We should also comment on the comparison to the finite-element method here; in the previous example comparison was to the symmetric collocation method for solving PDEs; in this case the comparison is more direct as the solution for the PDE in symmetric collocation is simply the posterior mean from the PMM. In this case we use a finite-element solver both to highlight the fact that the behaviour witnessed when using symmetric collocation is not unique to that solver, and because in existing methods for finding the multiple solutions to the Allen-Cahn equation the base numerical method applied is the finite-element method. Furthermore we note that as the underlying numerical method becomes arbitrarily accurate, the posterior inferences made in the inverse problem should be invariant to the forward solver used.

\section{Discussion} \label{sec:conclusion}

We have shown how to construct probabilistic models for the solution of partial differential equations, which quantify the uncertainty arising from numerical discretisation of the system. We have further shown how the uncertainty in the forward problem can be propagated into posteriors over parameters in inverse problems. This allows robust inferences to be made in inverse problems, even when the numerical scheme used to solve the forward problem is inaccurate, which is useful in cases where obtaining highly accurate solutions is computationally expensive, or where we are willing to tolerate less certain inferences in exchange for fast computation. In particular we have illustrated how this might be used to make inferences in nonlinear systems where a variety of phenomena, such as a non-unique solution could cause a numerical solver to fail.

Immediate extensions to this work lie in examining evolutionary systems in which the solution is additionally a function of time; the added complexity from the additional dimension demands more focussed attention. We also seek to examine a more generic approach for sampling from posterior distributions for nonlinear PDEs. Furthermore we note that the observations we have chosen for the forward problem are only one possible choice; another attractive option is given by Galerkin schemes for approximating PDEs, by choosing our observations to be Galerkin projections.

Lastly we seek to explore other choices of prior. The Gaussian measure is an unrealistic option in general, as it penalises extreme values and prevents encoding such simple properties as positivity of solutions.

\section{Acknowledgements}

TJS was supported by the Free University of Berlin within the Excellence Initiative of the German Research Foundation (DFG). MG was supported by EPSRC [EP/J016934/1, EP/K034154/1], an EPSRC Established Career Fellowship, the EU grant [EU/259348] and a Royal Society Wolfson Research Merit Award.

The authors would like to thank John Skilling for useful discussion, Patrick Farrell for providing code used in generating these results and Fran\c{c}ois-Xavier Briol for helpful feedback. In addition they express gratitude to the developers of the Python libraries Autograd and GPyOpt.

\bibliographystyle{plainnat}
\bibliography{jabref}

\begin{thebibliography}{11}
\providecommand{\natexlab}[1]{#1}
\providecommand{\url}[1]{\texttt{#1}}
\expandafter\ifx\csname urlstyle\endcsname\relax
  \providecommand{\doi}[1]{doi: #1}\else
  \providecommand{\doi}{doi: \begingroup \urlstyle{rm}\Url}\fi

\bibitem[Cialenco et~al.(2012)Cialenco, Fasshauer, and Ye]{Cialenco2011}
Igor Cialenco, Gregory~E Fasshauer, and Qi~Ye.
\newblock {Approximation of stochastic partial differential equations by a
  kernel-based collocation method}.
\newblock \emph{International Journal of Computer Mathematics}, 89\penalty0
  (18):\penalty0 2543--2561, December 2012.

\bibitem[Cockayne et~al.(2016)Cockayne, Oates, Sullivan, and
  Girolami]{Cockayne:2016ts}
Jon Cockayne, Chris Oates, Tim Sullivan, and Mark Girolami.
\newblock {Probabilistic Meshless Methods for Partial Differential Equations
  and Bayesian Inverse Problems}.
\newblock \emph{arXiv:1605.07811v1}, May 2016.

\bibitem[Conrad et~al.(2016)Conrad, Girolami, S{\"a}rkk{\"a}, Stuart, and
  Zygalakis]{Conrad:2016gv}
Patrick~R Conrad, Mark Girolami, Simo S{\"a}rkk{\"a}, Andrew Stuart, and
  Konstantinos Zygalakis.
\newblock {Statistical analysis of differential equations: introducing
  probability measures on numerical solutions}.
\newblock \emph{Statistics and Computing}, 2016.

\bibitem[Farrell et~al.(2015)Farrell, Birkisson, and Funke]{Funke2013}
Patrick~E Farrell, Asgeir Birkisson, and Simon~W Funke.
\newblock {Deflation techniques for finding distinct solutions of nonlinear
  partial differential equations}.
\newblock \emph{SIAM Journal on Scientific Computing}, 37\penalty0
  (4):\penalty0 A2026--A2045, 2015.

\bibitem[Fasshauer(1999)]{Fasshauer1999}
Gregory~E Fasshauer.
\newblock {Solving differential equations with radial basis functions:
  multilevel methods and smoothing}.
\newblock \emph{Advances in Computational Mathematics}, 11\penalty0
  (2-3):\penalty0 139--159, 1999.

\bibitem[Gelman(2006)]{Gelman2006}
A~Gelman.
\newblock {Prior distributions for variance parameters in hierarchical models
  (comment on article by Browne and Draper)}.
\newblock \emph{Bayesian analysis}, 1\penalty0 (3):\penalty0 515--534, 2006.

\bibitem[Hennig et~al.(2015)Hennig, Osborne, and Girolami]{Hennig:2015jf}
Philipp Hennig, Michael~A Osborne, and Mark Girolami.
\newblock {Probabilistic numerics and uncertainty in computations}.
\newblock \emph{Proc R Soc A}, 471\penalty0 (2179):\penalty0 20150142, July
  2015.

\bibitem[Owhadi(2015{\natexlab{a}})]{Owhadi2015}
Houman Owhadi.
\newblock {Bayesian numerical homogenization}.
\newblock \emph{Multiscale Modeling {\&} Simulation}, 13\penalty0 (3):\penalty0
  812--828, 2015{\natexlab{a}}.

\bibitem[Owhadi(2015{\natexlab{b}})]{Owhadi:2015vo}
Houman Owhadi.
\newblock {Multigrid with rough coefficients and multiresolution operator
  decomposition from Hierarchical Information Games}.
\newblock \emph{arXiv:1503.03467v4}, March 2015{\natexlab{b}}.

\bibitem[Stuart(2010)]{Stuart2010}
Andrew~M. Stuart.
\newblock Inverse problems: a {B}ayesian perspective.
\newblock \emph{Acta Numer.}, 19:\penalty0 451--559, 2010.
\newblock ISSN 0962-4929.

\bibitem[Wo{\'{z}}niakowski(2009)]{Wozniakowski:2009vj}
Henryk Wo{\'{z}}niakowski.
\newblock {What is information-based complexity?}
\newblock \emph{Essays on the complexity of continuous problems}, pages 89--95,
  2009.

\end{thebibliography}
\end{document}